\documentclass[prd,twocolumn,floats,floatfix,nofootinbib,superscriptaddress]{revtex4}
\usepackage{graphicx}
\usepackage{graphics}
\usepackage{dcolumn}
\usepackage{amssymb}
\usepackage{bm}
\usepackage{amsmath}
\usepackage{braket}
\usepackage[mathscr]{euscript}
\def\spose#1{\hbox to 0pt{#1\hss}}

\def\lta{\mathrel{\spose{\lower 3pt\hbox{$\mathchar"218$}}
     \raise 2.0pt\hbox{$\mathchar"13C$}}}
\def\gta{\mathrel{\spose{\lower 3pt\hbox{$\mathchar"218$}}
     \raise 2.0pt\hbox{$\mathchar"13E$}}}

\newcommand{\be}{\begin{equation}}
\newcommand{\en}{\end{equation}}
\newcommand{\bea}{\begin{eqnarray}}
\newcommand{\ena}{\end{eqnarray}}

\def\1{1 \!\! 1}

\newcommand{\determinante}{\mathop{\mathrm{Det}}}
\newcommand{\sgn}{\mathop{\mathrm{sgn}}}

\newcommand{\fresnels}{\mathop{\mathrm{S}}}
\newcommand{\fresnelc}{\mathop{\mathrm{C}}}
\newcommand{\erf}{\mathop{\mathrm{erf}}}

\begin{document}
\title{Consistent Probabilities in Perfect Fluid Quantum Universes}

\author{C.~R.~Bom}
\affiliation{ICRA - Centro Brasileiro de
Pesquisas F\'{\i}sicas -- CBPF, Rua Xavier Sigaud, 150, Urca,
CEP22290-180, Rio de Janeiro, Brazil}

\author{N.~Pinto-Neto}
\affiliation{ICRA - Centro Brasileiro de
Pesquisas F\'{\i}sicas -- CBPF, Rua Xavier Sigaud, 150, Urca,
CEP22290-180, Rio de Janeiro, Brazil}

\author{G.~B.~Santos}
\affiliation{ICRA - Centro Brasileiro de
Pesquisas F\'{\i}sicas -- CBPF, Rua Xavier Sigaud, 150, Urca,
CEP22290-180, Rio de Janeiro, Brazil}

\date{\today}

\begin{abstract}
Recently it has been claimed that the Wheeler-DeWitt quantization of gravity is unable to avoid cosmological singularities. However,
in order to make this assertion, one must specify the underlying interpretation of quantum mechanics which has been adopted. 
For instance, several nonsingular models were obtained in Wheeler-DeWitt quantum cosmology in the framework of the de Broglie-Bohm quantum theory. Conversely, there are specific situations where the singularity cannot be avoided in the framework of the Consistent Histories approach to quantum mechanics. In these specific situations, the matter content is described by a scalar field, and the Wheeler-DeWitt equation looks-like a Klein-Gordon equation.
The aim of this work is to study the Wheeler-DeWitt quantization of cosmological models where the matter content is described by an hydrodynamical perfect fluid, where the Wheeler-DeWitt equation reduces to a genuine Schr\"odinger equation. In this case, it is shown that the conclusions of the Consistent Histories and the de Broglie-Bohm approaches coincide in the quantum cosmological models where the curvature of the spatial sections is not positive definite, namely, that the cosmological singularities are eliminated. In the case of positive spatial curvature, the family of histories is no longer consistent and no conclusion can be given in this framework. 
\end{abstract}

\pacs{98.80.Cq, 04.60.Ds}

\maketitle

\section{Introduction}\label{intro}

The standard cosmological model has been successfully tested until the nucleosynthesis era, if one assumes the existence of the so called dark sector. However, the extrapolation of this scenario to higher energies leads to a singularity, where the volume of the universe becomes null, and physical quantities such as energy densities and the space-time curvature diverge in a finite cosmic time. A natural alternative is to investigate possible quantum effects in this high energy limit, which may avoid such cosmological singularities. Indeed, many investigations have shown the elimination of the singularities in quantum cosmological models using the Wheeler-DeWitt (WDW) \cite{deWitt} and Loop Quantum Gravity (LQG) quantization schemes \cite{ash4}. However, some authors have argued that in the WDW approach the singularity is not avoided \cite{CS,ash1,ash2,ash3,ash4}. 

In order to deal with these questions, one must define precisely how one can obtain physical information from the quantum state of the Universe: as we know, the standard Copenhagen interpretation cannot be used in quantum cosmology because it needs a classical external domain where measurements are made outside the quantum system in order to give a physical meaning to the quantum state, and in cosmology one is
dealing with a closed quantum system, the Universe, without any external classical observers. Hence, one has to use an alternative quantum theory, where external agents do not play any special role, and there are many viable proposals in the literature, for example the ones proposed in \cite{dBB, everett}. 

Among the papers mentioned above questioning the WDW approach, only two (\cite{ash4,CS}) have specified the interpretation of quantum mechanics they are using, which was the Consitent Histories approach \cite{griffiths84, omnes}. 
In this framework, probabilities are not assigned
to events as in usual quantum mechanics, but to whole histories.
A history of a closed physical system is a succession of properties
of this system occurring at different times.
An example of a property of a system is the sentence `the eigenvalue of
the observable $\hat{B}$ is in the set $D$'. To each property is associated
a projector operator. In the above example, it would be the projector $\hat P$
onto the subspace of the Hilbert space containing all eigenvectors of $\hat{B}$
with eigenvalues in the set $D$. 
However, as we know, we cannot assign probabilities
to every history in quantum mechanics. The interference figure
obtained from the two slit experiment is an evidence of this
fact. Hence, we must establish
what are the conditions on families of histories in order to be
possible to assign probabilities to all members of such families.
Once we obtain these conditions, we will have the possibility
of saying that a history of the Universe is more probable than another one,
without mentioning observers or measurements.

The Wheeler-DeWitt quantization of a spatially flat Friedmann cosmological model with a massless free scalar field
was implemented under this framework. The resulting Wheeler-DeWitt equation is a Klein-Gordon like equation.
In that case, as shown by Craig and Singh \cite{CS}, the cosmological singularities are not avoided
if one takes the square-root of the Klein-Gordon equation and their associated Newton-Wigner states, and histories concerning properties of the quantum system at only two specific moments of time: the infinity past and the infinity future. However, as shown in Ref.~\cite{nelsonetal}, if one considers the two frequency sectors of the resulting Klein-Gordon like equation and/or three or more instants of time, the family of histories are not consistent anymore, and no conclusion about the existence of singularities can be made. Furthermore, if one analyzes the same physical system within the de Broglie-Bohm quantum theory,
in which trajectories can be unambiguosly defined through the so called guidance relations, there are plenty of states where quantum bounces occurs in both quantization schemes, and the cosmological singularity disappears. Also, in Ref.~\cite{nelsonetal} the reasons for these discrepant results coming from these two different quantum approaches were discussed: it is due to the fact that the resulting Wheeler-DeWitt equation of the minisuperspace quantum cosmological model considered is a Klein-Gordon like equation, not an usual Schr\"odinger like equation. As we know, the predictions of these two approaches agree in usual physical systems investigated in the laboratory, which obey the Schr\"odinger equation.

The aim of this paper is to discuss carefully this same question concerning the existence of singularities in Friedmann-Lema\^itre-Robertson-Walker (FLRW) quantum universes, but now considering that the matter content is described by a perfect hydrodynamical fluid\footnote{We will choose this fluid to be radiation, because at high energies the rest energy of any particle becomes negligible with respect to their total kinetic energies. Furthermore, other possibilities will lead essentially to the same results.}, in which case it is known that the resulting Wheeler-DeWitt equation is a Schr\"odinger like equation.
In this situation, we expect that the Consistent Histories approach should give the same results as the ones already obtained within the de Broglie-Bohm quantum theory, namely, that the cosmological singularities are eliminated \cite{nelson1,nelson3, NSF,nelson2}. And indeed, our analysis have shown that in the cases where it is possible to define consistent probabilities for histories (the models where the spatial curvature is not positive definite) evaluated in two specific moments ($t_1 \rightarrow - \infty$ and $t_2 \rightarrow + \infty$), the histories without singularities have probability one to occur, as in the de Broglie-Bohm case. For positive spatial curvature, we have shown that no consistent probability can be associated to histories concerning those two moments.   

The paper will be divided as follows: in section \ref{CH} we present the Consistent Histories framework. In subsection \ref{classico} we present the classical cosmological model and in subsection \ref{quantico} we discuss the quantization of the latter. In section \ref{CSFLRW} we discuss the FLRW model in the Consistent Histories interpretation and then, in subsection \ref{singularity_criteria}, we present the criterion for a nonsingular universe in this context. In subection 
\ref{resultados} we evaluate the probabilities of the FLRW quantum universes to reach a singularity in $t \rightarrow \pm \infty$ in flat and open universes and in \ref{fechado} we discuss the closed case. Finally in \ref{conclusao} we present our final remarks. Numerical calculations supporting our results are shown in the appendix.

\section{Consistent Histories}\label{CH}

The Consistent Histories approach, also known as Generalized Quantum Mechanics (GQM), was proposed initially by Griffiths \cite{griffiths84} and Omnes \cite{omnes}, and later developed in the gravity context mainly by Hartle and Gell-Mann \cite{gell-mann, hartle} and Halliwell \cite{halliwell91, halli99, halli01}. It is a framework in which probabilistic predictions can be made about quantum systems avoiding the need of external classical observers, and preserving most of the conceptual structure of the standard Copenhagen interpretation. In this section, we will present some general aspects of this interpretation that are relevant to the issue of the presence of singularities in quantum cosmology.
In this approach, instead of associating probabilities to a single eigenvalue, we assign them to a history, that is, a sequence of time ordered eigenvalue intervals. For a given set of observables $A^\alpha$ with eigenvalue intervals $\Delta a^\alpha_k$, a history $h$ may be represented by a class operator $C_{h}$ defined as:

\begin{equation}
C_h=P_{\Delta a_{k_1}}^{\alpha_1}(t_1)... P_{\Delta a_{k_{n-1}}}^{\alpha_{n-1}}(t_{n-1})P_{\Delta a_{k_{n}}}^{\alpha_{n}}(t_n), 
\label{operador_historia}
\end{equation}
where $P_{\Delta a_{k}}^{\alpha}(t)$ is the projector onto the subspace for which the $k$th eigenvalue of the observable $A^{\alpha}$ in time $t$ is in the interval $\Delta a_{k}$. The time dependent projectors are defined using the Heisenberg operators,
\begin{equation}
P_{\Delta a_{k}}^{\alpha}(t)= U^\dagger(t)P_{\Delta a_k}^{\alpha}U(t),
\end{equation}
where $U(t)$ is the evolution operator of the quantum system with Hamiltonian $H$,

\begin{equation}
U(t_i-t_j)=\exp{\left [-\imath H(t_i-t_j) \right ]},
\end{equation}
where we considered $\hbar=1$ and $\imath$ is the imaginary unity. 

We are interested in asking questions about the eigenvalue intervals of the scale factor operator, $\hat{R}$, at some moments of time. The time independent projector operator associated with this is

\begin{equation}
P_{\Delta R}= \int_{\Delta R} dR \Ket{R}\Bra{R} .  
\label{projetor_usual}
\end{equation}

At this point, we may distinguish two kinds of histories. The fine-grained histories are the most precise descriptions of a physical system, and deals with the smallest intervals of eigenvalues. On the other hand, the coarse-grained histories are partitions of the fine-grained ones, considering larger intervals of eigenvalues. Those histories are defined to answer physical relevant questions, which in our case is whether the universe is singular or not.

The wave function that represents the amplitude for a given initial state to follow the history $h$, the branch wave function, is defined as

\begin{equation}
\Ket{\Psi_h}=C_h^\dagger\Ket{\Psi},
\label{funcao_onda_historia}
\end{equation} 
which is not normalized. The branch wave function in (\ref{funcao_onda_historia}) represents a physical system evolving through a set of instants of time $t_i$, generally discontinuous. Since it is not a function of time $t$, the branch wave function does not satisfy the Schr\"odinger equation. 
 
In order to have a meaningful notion of probability, we must require some reasonable conditions upon the probability $p$ we want to assign to each history:

\begin{subequations} \label{cond}

\begin{equation}
0 \leq p (C_h) \leq 1,
\label{cond_1}
\end{equation}

\begin{equation}
p(\1)=1,
\label{cond_2}
\end{equation}

\begin{equation}
p(C_h+C_{h^\prime})= p(C_h) + p(C_{h^\prime}),
\label{cond_3}
\end{equation}
\end{subequations}
where $\1$ is the identity operator. Following the Hartle-Gell-Mann \cite{gell-mann} approach, we define the decoherence functional for a set of histories $\{ h_i \}$, considering pure states only, as

\begin{equation}
D(h,h')= \Braket{\Psi_{h} | \Psi_{h'}}.
\label{funcional_descoerencia}
\end{equation}

It has been shown that in order to satisfy (\ref{cond}), the sufficient condition is:

\begin{equation}
D(h,h')= 0, \quad h \neq h'.
\label{consistencia}
\end{equation}
If the family of histories satisfy (\ref{consistencia}), we say that this family is consistent.
Since $\sum_k P_{\Delta a_k}^{\alpha}=1$, we have
\begin{eqnarray}
\sum_h C_h & = & \sum_{k_1}\sum_{k_2}...\sum_{k_n}P_{\Delta a_{k_1}}^{\alpha_1}(t_1) P_{\Delta a_{k_2}}^{\alpha_2}(t_2)...P_{\Delta a_{k_n}}^{\alpha_n}(t_n) \nonumber \\
&=&1 \quad,
\end{eqnarray}
and the decoherence functional is normalized because
\begin{equation}
\sum_h \Ket{\Psi_h} = \sum_h C_h^\dagger \Ket{\Psi}= \Ket{\Psi}.
\label{soma_historias_geral}
\end{equation}
The probability of a history can then be written as

\begin{eqnarray}
p&=&|P_{\Delta a_{k_1}}^{\alpha_1}(t_1)... P_{\Delta a_{k_{n-1}}}^{\alpha_{n-1}}(t_{n-1})P_{\Delta a_{k_{n}}}^{\alpha_{n}}(t_n) \Ket{\Psi}|^2 \nonumber \\
&=&  |C_h^\dagger \Ket{\Psi}|^2=\Braket{\Psi_h|\Psi_h}. 
\label{probabilidade_hist}
\end{eqnarray}
Using (\ref{consistencia}), (\ref{soma_historias_geral}) and (\ref{probabilidade_hist}), and considering a consistent family of histories, we can write the decoherence functional as

\begin{equation}
 D(h,h') = \Braket{\Psi_h|\Psi_{h'}} = p(h)\delta_{h'h},
\label{funcional_consistencia_diagonalizado}
\end{equation}
where $p(h)\equiv p(C_h)$ represents the probability that the system follows a history $h$. One can interpret condition (\ref{consistencia}) as demanding that no interference can occur between histories.

%%%%%%%%%%%%%%%%%%%%%%%%%%%%%% 
\section{Radiation dominated FLRW}\label{model}

With the quantum interpretation specified in the last section, we will calculate the probabilistic predictions concerning the existence of singularities in specific cosmolgical models. In this section we will introduce FLRW models where the matter content is described by a radiation fluid with $p=\rho/3$ \cite{nivaldo96, barros98}. The result can easily be generalized to any equation of state $p=w\rho$ \cite{alvarenga, peter06}. 

\subsection{Classical model}\label{classico}
The line element of the homogeneous and isotropic FLRW geometries can be written as

\begin{equation}
ds^2=g_{\mu \nu}dx^\mu dx^\nu=-N^2(t)dt +R^2(t)\sigma_{ij}dx^idx^j,
\label{metrica}
\end{equation}
where $N(t)$ is the lapse function, $R(t)$ is the scale factor and $R^2\sigma_{ij}$ is the metric of the homogeneous and isotropic 3-space of constant curvature $k=-1,0,1$, corresponding to the hyperbolic, flat and spherical geometries, respectively. We also set $c=1$. The latin indexes vary from 1 to 3 and the greek indexes vary from 0 to 3. 
The matter content is a perfect fluid, where we use the Schutz \cite{schutz} formalism to relativistic fluids. 
The total action reads,

\begin{eqnarray}
S &=& \int d^4x \mathscr{L} = \int d^4x \mathscr{L}_g+\mathscr{L}_f \nonumber \\
&=& \int_M d^4x\sqrt{-g} {}^4R   + 2\int_{\partial M^4}d^3x\sqrt{-h}  h_{ij}K^{ij}  \nonumber \\ 
&& + \int_M d^4x \sqrt{-g}\,p,
\label{acao}
\end{eqnarray}
where the last term corresponds to the fluid action, $p$ is the pressure, $g=\determinante (g_{\mu \nu} )$, $K_{ij}$ is the extrinsic curvature, $h_{ij}$ is the $3$-metric, $h=\determinante (h_{ij})$, $\partial M^4$ is the boundary of the manifold $M^4$ and ${}^4R $ is the Ricci curvature escalar constructed with the metric $g_{\mu \nu}$. 

For the geometry represented by (\ref{metrica}) and fluid described by radiation, one can show that the above action (\ref{acao}) 
reduces to \cite{lap_ruba, nivaldo96,barros98}

\begin{equation}
S= \int dt (p_R\dot{R}+p_{T}\dot{T}-N\mathscr{H}),
\label{acao_hamiltoniana}
\end{equation}
where $T$ is the degree of freedom associated to the radiation fluid, $p_R$ and $p_T$ are the momenta canonically 
conjugate to $R$ and $T$, respectively, and it can be seen that the lapse function $N(t)$ is the lagrange multiplier of the super-hamiltonian 

\begin{equation}
\mathscr{H}=-\frac{p_R^2}{24R}-6kR+\frac{p_T}{R}.
\end{equation}
The momentum  $p_T$ is linear in the super hamiltonian, so we choose the time gauge $N=R$ (conformal time), which
reduces the action to its canonical form
\begin{equation}
S= \int dt (p_R\dot{R}+\frac{p_R^2}{24}+6kR^2),
\end{equation}
where one can read the canonical hamiltonian $$H\equiv-\frac{p_R^2}{24}-6kR^2,$$ and that $t$ is the conformal time.  

The equations of motion are
\begin{equation}
\frac{dR}{dt}=-\frac{p_R}{12}, \qquad \frac{dp_R}{dt}=12kR.
\end{equation}
The differential equation for $R$ can be written as
\begin{equation}
\ddot{R}+kR=0,
\end{equation}
which has the following solutions
\begin{equation}
R(t)
=R_0
\left\{
\begin{array}{lcl}   
         \sinh{t}      && k=-1 ,  \\
         t             && k=0   , \\
         \sin{t}       && k=+1   .                  
\end{array}\right.  
\label{eq:classol}
\end{equation}
In all cases the universe reaches an unavoidable classical singularity in $t=0$. In the next subsection we present the quantum model for the radiation dominated FLRW universe in order to discuss whether the singularity persists in the quantum regime.

\subsection{Canonical quantization and boundary conditions }\label{quantico}

As we have reduced the action to its canonical form, the canonical quantization will lead to a Schr\"odinger like equation,
\begin{equation}
\label{schpsi2}
i\frac{\partial\Psi}{\partial t}=\hat{H} \Psi.
\end{equation}

The above Schr\"odinger equation is simple to solve, but one must take care of the fact that $R$ is defined only for $R \geq 0$. As a result, the description of the quantum system requires not only the canonical quantization process but also the imposition of boundary conditions to ensure that the wave functions $\Psi$ are square integrable, and the hamiltonian operator is self-adjoint. The self adjointness of the hamiltonian is a necessary condition to have unitary evolution. We may construct the hamiltonian operator by performing the substitution $p_R \rightarrow -\imath \frac{d}{dR}$ yielding
\begin{equation}
\hat{H}=\frac{1}{24}\frac{d^2}{dR^2}-6kR^2,
\label{operador_hamiltoniano}
\end{equation}
where $\hbar=1$. The self adjointeness condition may be expressed as
\begin{equation}
\Braket{\Psi_1|\hat{H}\Psi_2}=\Braket{\hat{H}\Psi_1|\Psi_2}.
\label{condSA}
\end{equation}
From this point, in order to simplify the notation, we will omit the ($\hat{}$). The condition (\ref{condSA}) is equivalent to
\begin{equation}
\int_0^\infty dR \Psi^\star(R)\frac{d^2 \Psi(R)}{dR^2} = \int_0^\infty dR \frac{d^2\Psi^\star(R)}{dR^2} \Psi(R).
\end{equation}
After integration by parts twice we get
\begin{align}
& \left (\Psi^\star_1(R)\frac{d \Psi_2(R) }{dR} - \frac{d \Psi^\star_1(R) }{dR} \Psi_2(R) \right )(+\infty) \nonumber  \\
&= \left (\Psi^\star_1(R)\frac{d \Psi_2(R) }{dR} - \frac{d \Psi^\star_1(R) }{dR} \Psi_2(R) \right)(0).
\label{condicao_contorno_parcial}
\end{align}
As we want to find only square integrable solutions, the left side of the equation (\ref{condicao_contorno_parcial}) must vanish, so
\begin{equation}
\left (\Psi^\star_1(R)\frac{d \Psi_2(R) }{dR} - \frac{d \Psi^\star_1(R) }{dR} \Psi_2(R) \right )(0)=0.
\label{condicao_contorno_parcial2}
\end{equation}
The necessary and sufficient condition for (\ref{condicao_contorno_parcial2}) to be satisfied is
\begin{equation}
\Psi'(0)=\eta\Psi(0),
\label{condicao_contorno_parcial3}
\end{equation}
where $\eta$ is in the interval $(-\infty,+\infty)$ and $'$ denotes derivative with respect to $R$. For simplicity we restrict ourselves to the case $\eta=0$, using the explicit results presented in \cite{nivaldo96}.

Let $G(R,R',t)$  be the propagator of the hamiltonian (\ref{operador_hamiltoniano}) in the Hilbert space $L^2(-\infty,+\infty)$. The propagator $G^{(a)}(R,R',t)$ which satisfies $\Psi'(0)=0$  in the half line Hilbert space $L^2(0,+\infty)$ is
\begin{equation}
G^{(a)}(R,R',t)=G(R,R',t)+G(R,-R',t),
\label{propagadorA}
\end{equation}
where $G(R,R',t)$ is the usual quantum harmonic oscillator propagator for which the mass is $m=12$ and the frequency is $\omega=\sqrt{k}$ \cite{feynman}. It can be written as
\begin{align}
G(R,R',t) &= \left ( \frac{6\sqrt{k}}{\pi \imath \sin{\sqrt{k}t}} \right )^{1/2} \nonumber \\
 &  \exp \left \{ \frac{6\imath \sqrt{k}}{\sin{\sqrt{k}t}} \left ( ( R^2+R'^2  ) \cos{\sqrt{k}t} - 2RR' \right ) \right \} 
\end{align}

\subsection{Dynamics of the radiation dominated FLRW quantum models}\label{dinamica_quantica}
In order to analyze the dynamical evolution of this model, we choose as initial condition the wave function
\begin{equation}
\Psi_0^{(a)} (R)=\sqrt[4]{\frac{8\sigma}{\pi}}\exp(-\beta R^2), 
\end{equation}
where $\sigma \in \mathbb{R}$ , $\sigma > 0$ e $\beta=\sigma+\imath p$. Using the propagator defined in (\ref{propagadorA}), we evaluate $\Psi (R,t)$
\begin{eqnarray}
\Psi (R,t) &=& \int_0^\infty dR \cdot G^{(a)}(R,R',t)\Psi_0^{(a)} (R) \nonumber \\ 
&=& \int_{-\infty}^{+\infty}  dR \cdot G(R,R',t)\Psi_0^{(a)} (R), 
\label{sol_wf}
\end{eqnarray}
which results
\begin{widetext}
\begin{align}
 \Psi (R,t)= &  \nonumber \\
 & \left \{ \frac{12\sqrt{2\sigma} \sqrt{k}}{\sqrt{\pi}\cos{(\sqrt{k}t)}(\beta \tan{(\sqrt{k}t)} -6\imath\sqrt{k})} \right \}^{1/2}  \exp \left \{ R^2  \frac{6\imath \sqrt{k}}{\tan{(\sqrt{k}t)}} \left ( 1+ \frac{6\imath \sqrt{k}}{\cos^2{(\sqrt{k}t)} \left ( \beta \tan{(\sqrt{k}t)} - 6\imath \sqrt{k} \right )} \right) \right \}. 
\label{funcao_de_onda} 
\end{align}
\end{widetext}
from which we calculate the mean values
\begin{equation}
\Braket{R(t)} =  \kappa \left \{\sigma^2\sin^2{\sqrt{k}t} + (6-p\tanh{\sqrt{k}t})^2\cosh^2{\sqrt{k}t}, \right \}^{1/2}
\label{eq:classol}
\end{equation}
where $\kappa := \frac{\sqrt{2}}{12(\pi \sigma)^{1/2}}$. In all cases $k=-1,0,+1$ the mean values of $R$ never reach the singular value $R=0$. However, this is not a sufficient condition to state that the present model is never singular. In the next section we present the consistent histories approach to deal with the singularity issue in this context.

%%%%%%%%%%%%%%%%%%%%%%%%%%%%%%%%
\section{Consistent probabilities in radiation dominated FLRW quantum models}\label{CSFLRW}

In this section we present a precise criterion to determine if a given quantum model can avoid the singularity. This criterion is developed in \cite{CS}. In the subsection (\ref{singularity_criteria}) we specify this criterion and perform some calculations in order to simplify and make the problem more tractable. After that, we evaluate the quantum model presented in section \ref{quantico}. 

\subsection{The singularity in Consistent Histories \label{singularity_criteria}} 
 
To address the question whether the universe is singular or not, one must construct a suitable family of histories. To pursue that, we focus on the behavior of the scale factor. We say that a history is singular if the scale factor is within an interval $\Delta R_\star= [0,R_\star]$, where $R_\star$ is a fiducial value, which can be arbitrarily small. This interval represents the universe arbitrarily near the singularity. We also define the complementary interval   $\overline{\Delta R_\star}= (R_\star,+\infty)$, where the universe is out of the singularity. The class operator that represents the history without singularity is
\begin{equation}
C_b(t_1,...,t_n)= P_{\overline{\Delta R_\star}}(t_n) P_{\overline{\Delta R_\star}}(t_{n-1}) ... P_{\overline{\Delta R_\star}}(t_1),
\end{equation}
for $n$ moments in time, where the $P_{\overline{\Delta R_\star}}(t)$ are the projectors in the interval $\overline{\Delta R_\star}$. Any history for which at least one projector defined in the singularity interval gives a nonnull value when applied to the state of the system is a singular history. From now on, for simplicity, we consider histories specifying two moments $t_1$ and $t_2$ evaluated in the limits $t_1 \rightarrow -\infty$ and $t_2 \rightarrow +\infty$. The class operator for the singularity is
\begin{equation}
 C_{s}(t_1,t_2) = P^{1}_{\Delta R_\star} P^{2}_{\overline{\Delta R_\star}} +  P^{1}_{\overline{\Delta R_\star}} P^{2}_{\Delta R_\star} + P^{1}_{\Delta R_\star} P^{2}_{\Delta R_\star}, 
\label{defsingular} 
\end{equation}      
with $P^{1}_{\Delta R}:= P_{\Delta R}(t_1)$ and $P^{2}_{\Delta R}:= P_{\Delta R}(t_2)$. The class operator in (\ref{defsingular}) can be more conveniently written as \cite{CS}
\begin{equation}
C_{s}(t_1,t_2) = P^{1}_{\Delta R_\star} + P^{2}_{\Delta R_\star} - P^{1}_{\Delta R_\star} P^{2}_{\Delta R_\star}
\end{equation}
where we considered $P^{1}_{\Delta R_\star} P^{2}_{{\Delta R_\star}} + P^{1}_{\Delta R_\star} P^{2}_{\overline{\Delta R_\star}} = P^{1}_{\Delta R_\star}$. 
The criterion for consistency of the family is 
\begin{align}
\Bra{\Psi} C_{b} C_{s}^{\dagger}\Ket{\Psi} &=  \Bra{\Psi}P^{1}_{\overline{\Delta R_\star}}P^{2}_{\overline{\Delta R_\star}}P^{1}_{\Delta R_\star}\Ket{\Psi} \nonumber \\
&\quad + \Bra{\Psi}P^{1}_{\overline{\Delta R_\star}}P^{2}_{\overline{\Delta R_\star}}P^{2}_{\Delta R_\star}\Ket{\Psi} \nonumber \\
& \quad - \Bra{\Psi}P^{1}_{\overline{\Delta R_\star}}P^{2}_{\overline{\Delta R_\star}}P^{2}_{\Delta R_\star}P^{1}_{\Delta R_\star}\Ket{\Psi}=0.
\label{consistencia_completaeq}
\end{align}
Nevertheless, the last two terms are null
since they contain products of projectors in orthogonal subspaces. The relation for the consistency of the family of histories may also be written in terms of projectors in the singularity interval only:
\begin{align}
\Bra{\Psi}(1- C_{s}) C_{s}^{\dagger}\Ket{\Psi} &= \Bra{\Psi} C_{s}^{\dagger}\Ket{\Psi} - \Bra{\Psi} C_{s} C_{s}^{\dagger}\Ket{\Psi} \nonumber \\
&= \Braket{\Psi | \Psi_s} - \Braket{\Psi_s | \Psi_s} \label{consistencia_singular} \nonumber\\
&= \Braket{\Psi |P^{2}_{\Delta R_\star}P^{1}_{\Delta R_\star}|\Psi} \nonumber \\
& \quad - \Braket{\Psi |P^{1}_{\Delta R_\star}P^{2}_{\Delta R_\star}P^{1}_{\Delta R_\star}|\Psi} ,
\end{align}
where we used that $C_{s}+C_{b}=1$. 

If the histories decohere, the probability for having the singularity is
\begin{align}
p_s\equiv\Bra{\Psi} C_{s} C_{s}^{\dagger}\Ket{\Psi} &= - \Bra{\Psi}P^{1}_{\Delta R_\star}P^{2}_{\Delta R_\star}P^{1}_{\Delta R_\star}\Ket{\Psi}  \nonumber \\
& \quad  +\Bra{\Psi}P^{2}_{\Delta R_\star}\Ket{\Psi} +  \Bra{\Psi}P^{1}_{\Delta R_\star}\Ket{\Psi}.
\label{prob_singularidade}
\end{align}

The probability that the universe is not singular is given by
\begin{align}
p_b\equiv\Bra{\Psi}(1- C_{s})(1- C_{s}^{\dagger})\Ket{\Psi} &= 1-2\mathrm{Re}(\Braket{\Psi | \Psi_s}) + \Braket{\Psi_s | \Psi_s},
\label{prob_ricochete}
\end{align} 
where we considered again that $C_{s}+C_{b}=1$. In the next section we evaluate these probabilities and check whether the histories decohere in the FLRW quantum model filled with radiation.

\subsection{Singularity in radiation dominated FLRW models} \label{resultados}

Consider the wave function in (\ref{sol_wf}):
\begin{align}
\Braket{R|\Psi(t)}=\Psi (R,t) &=  \int_{-\infty}^{+\infty}  dR \cdot G(R,R',t)\Psi_0^{(a)} (R), 
\label{sol_wf2}
\end{align}
which is square integrable for a given initial condition $\Psi_0^{(a)} (R)=\Braket{R|\Psi}$. If we apply the projector in the singularity interval to a state $\Ket{\Psi}$ we get

\begin{equation}
\Ket{\Phi(t)}\equiv P_{\Delta R_\star}(t)\Ket{\Psi} = U^{\dagger}(t)\int^{R_{\star}}_{0} dR \Ket{R} U(t) \Braket{R|\Psi}. 
\label{wf_projb}
\end{equation}
Inserting the resolution of the identity $\int^{+\infty}_{0}dR' \Ket{R'}\Bra{R'}$ in the equation above yields 
\begin{align}
\Ket{\Phi(t)}=\int^{+\infty}_{0}dR' \Ket{R'}\Bra{R'} U^{\dagger}\int^{R_{\star}}_{0} dR \Ket{R} \Braket{R|\Psi(t)} \nonumber \\
 = \quad\int^{+\infty}_{0}dR' \Ket{R'} \int^{R_{\star}}_{0} G^{(a)}(R,R',-t) \Psi(R,t)dR
\label{wf_proj2}
\end{align}
Let us evaluate the integral
\begin{eqnarray}
\int^{R_{\star}}_{0} G^{(a)}(R,R',-t) \Psi(R,t)dR
\label{projected_wf}
\end{eqnarray}
in the limits $t \rightarrow  \pm \infty$ by considering its square modulus. It follows from the Cauchy-Schwarz inequality that

\begin{align}
\int^{R_{\star}}_{0} |G^{(a)}(R,R',-t) \Psi(R,t)|^2 dR & \leq \nonumber \\
\left ( \int^{R_{\star}}_{0}   |G^{(a)}(R,R',-t)|^2\,dR \right )& \times \left(\int^{R_{\star}}_{0} |\Psi(R,t)|^2dR \right ).
\label{projected_wf_a}
\end{align}

The last integral in the inequality above is a real and positive number $c^2$ with $ 0 \leq c^2 \leq 1$ for a given time $t$. Let us analyze the square modulus of $G^{(a)}(R,R',-t)$, which yields\footnote{Since we are interested in the limits $t \rightarrow \pm \infty$ we will ignore the minus sign in $t$.}

\begin{align}
|G^{(a)}(R,R',t)|^2 &= |G(R,R',t) + G(R,-R',t)|^2 \nonumber \\
&=  \frac{12 \sqrt{|k|}}{\pi |\sin(\sqrt{k}t)|} \left ( 1+ \cos \left ( \frac{24\sqrt{k}RR'}{\sin(\sqrt{k}t)} \right) \right )
\label{g_modulus}
\end{align}
Note that the first term in (\ref{g_modulus}) does not depend on $R$. We may rewrite the integral of (\ref{g_modulus}) as

\begin{align}
 \frac{12 \sqrt{|k|}}{\pi |\sin(\sqrt{k}t)|} \int^{R_{\star}}_{0} dR \left ( 1+ \cos \left ( \frac{24\sqrt{k}RR'}{\sin(\sqrt{k}t)} \right) \right ) = \nonumber \\
  \frac{12}{\pi | \sin(\sqrt{k}t)|} \Bigg |^{R^{\star}}_{0}  +  \frac{\sin(\sqrt{k}t)}{2 R' |\sin{\sqrt{k}t}|} \int^{u^{\star}}_{0} \cos(u) du  =  \nonumber \\
\frac{12 R^{\star}}{\pi |\sin{(\sqrt{k}t)}|} + \frac{1}{2R'} \sin \left ( \frac{ 24 R' R^{\star}}{\sin( \sqrt{k}t)} \right ) \quad ,   
\label{g_modulus_int}
\end{align}
where we defined $u=24RR'/\sin(\sqrt{k}t)$ and $u^{\star}=24R^{\star}R'/\sin(\sqrt{k}t)$. In the spatially flat and hyperbolic universes in the limits $t \rightarrow \pm \infty $ we have $ u^{\star} \rightarrow 0$.

Using Eq.\ (\ref{g_modulus_int}), we see that the limits considered for the integral of $|G^{(a)}(R,R',t)|^2$ in the spatially flat and hyperbolic universes are

\begin{eqnarray}
\lim_{t\rightarrow\pm\infty}\int^{R^{\star}}_0 |G^{(a)}(R,R',-t)|^2 dR=
\left\{
\begin{array}{lcl}   
         0   ,          && k=-1,   \\
         0   ,          && k=0.                          
\end{array}\right.
\label{limit_G}
\end{eqnarray}

As $\int^{R_{\star}}_{0} |\Psi(R,t)|^2dR$ is limited in the interval $[0,1]$ the right side of Eq.\ (\ref{projected_wf_a}) is null, hence we have
\begin{eqnarray}
\lim_{t\rightarrow\pm\infty} \left|\int^{R_{\star}}_{0} G^{(a)}(R,R',-t) \Psi(R,t)dR\right| = 0,
\end{eqnarray}
for $k=-1$ and $k=0$. Therefore, we have shown that $P_{\Delta R_\star}(t)\Ket{\Psi}$ vanishes in the infinite past and in the infinite future whatever the initial state.
According to equations (\ref{consistencia_singular}), (\ref{prob_singularidade}) and (\ref{prob_ricochete}) we then have
\begin{subequations}
\begin{equation}
\lim_{\substack{t_1\rightarrow -\infty\\ t_2\rightarrow +\infty }} \Braket{\Psi_b|\Psi_s}=0 ,
\end{equation}
\begin{equation}
\lim_{\substack{t_1\rightarrow -\infty\\ t_2\rightarrow +\infty }} \Braket{\Psi_s|\Psi_s}=0 ,
\end{equation}
\begin{equation}
\lim_{\substack{t_1\rightarrow -\infty\\ t_2\rightarrow +\infty }} \Braket{\Psi_b|\Psi_b}=1 .
\end{equation}
\end{subequations}
That is, the set of histories is consistent and the probability of having a singularity is null for the flat and open Friedmann universes for any initial state.

In order to check this result explicitly, let us consider the initial condition $\Psi_0^{(a)} (R)=\sqrt[4]{\frac{8\sigma}{\pi}}\exp(-\sigma R^2)$, with $\sigma >0$. The projected state in the singularity interval is

\begin{eqnarray}
P_{\Delta R_\star}(t)\Ket{\Psi}&=&U^{\dagger}(t)\int^{R_{\star}}_{0} dR \Ket{R} U(t) \Braket{R|\Psi} 
\label{wf_proj}
\end{eqnarray}
As before, inserting the identity $\int^{+\infty}_{0}dR' \Ket{R'}\Bra{R'}$, the equation above takes the form 
\begin{eqnarray}
\int^{+\infty}_{0}dR' \Ket{R'}\Bra{R'} U^{\dagger}(t)\int^{R_{\star}}_{0} dR \Ket{R} \Braket{R|\Psi(t)} \nonumber \\
= \int^{+\infty}_{0}dR' \Ket{R'} \int^{R_{\star}}_{0} G^{(a)}(R,R',-t) \Psi(R,t)dR.
\label{wf_proj2v2}
\end{eqnarray}
We evaluate the following integral
\begin{eqnarray}
I = \int^{R_{\star}}_{0} G^{(a)}(R,R',-t) \Psi(R,t)dR,
\label{project_wfv2}
\end{eqnarray}
which results for the $k \rightarrow 0$ case in
\begin{align}
I &=  \frac{(2\pi)^{1/4}}{\sigma^{1/4}\exp{(-\sigma R^{\prime 2})}} \nonumber \\
& \quad \times  \left[ \erf{\left (\frac{6\imath (R_{\star} - R^\prime) + \sigma R^\prime t}{\sqrt{\imath 6 t -\sigma t^2}} \right )  }\right. \nonumber \\ 
& \left.\qquad + \erf{\left ( \frac{6\imath (R_{\star} + R^\prime) - \sigma R^\prime t}{\sqrt{\imath 6 t -\sigma t^2}} \right ) }  \right] .
\end{align}
In the considered limits $I$ vanishes, hence we have
\begin{equation}
\lim_{\substack{t\rightarrow \pm \infty }} P_{\Delta R_\star}(t)\Ket{\Psi}=0, 
\end{equation}
and the family of histories is consistent since
\begin{equation}
\Bra{\Psi} C_{b} C_{s}^{\dagger}\Ket{\Psi} = 
 \Bra{\Psi}P^{1}_{\overline{\Delta R_\star}}P^{2}_{\overline{\Delta R_\star}}P^{1}_{\Delta R_\star}\Ket{\Psi} =0.
\end{equation}

%The projection in the interval $\overline{\Delta R}= [R_\star,+\infty]$ with null curvature is
%\begin{eqnarray}
%\lim_{\substack{t\rightarrow \pm \infty }} \phi_{\overline{\Delta R_\star}} = \frac{(2\pi)^{1/4}}{\sigma^{1/4}\exp{-\sigma R^{\prime 2}}} \nonumber \\
%&\times& \erf{\left (\frac{6 +\imath \sigma R}{\sqrt{\sigma }} \right )  } - \erf{\left ( \frac{ -6 + +\imath \sigma R}{\sqrt{ -\sigma }} \right ) } 
%\end{eqnarray}
%\begin{eqnarray}
% \phi_{\overline{\Delta R_\star}} = \frac{(2\pi)^{1/4}}{\sigma^{1/4}\exp{-\sigma R^{\prime 2}}} \nonumber \\
%&\times& \erf{\left (\frac{1/2(1+\imath)(6-6\imath +\imath \sigma R+ +\sigma R)}{\sqrt{\sigma }} \right )  } - \erf{\left ( \frac{1/2(1+\imath)( \imath6 -6 \sigma R +\imath \sigma R)}{\sqrt{ -\sigma }} \right ) } 
%\end{eqnarray}
According to equations (\ref{prob_singularidade}) and (\ref{prob_ricochete}) we have
\begin{subequations}
%\begin{equation}
%\lim_{\substack{t_1\rightarrow -\infty\\ t_2\rightarrow +\infty }} \Braket{\Psi_b|\Psi_s}=0 ,
%\end{equation}
\begin{equation}
\lim_{\substack{t_1\rightarrow -\infty\\ t_2\rightarrow +\infty }} \Braket{\Psi_s|\Psi_s}=0, 
\end{equation}
\begin{equation}
\lim_{\substack{t_1\rightarrow -\infty\\ t_2\rightarrow +\infty }} \Braket{\Psi_b|\Psi_b}=1. 
\end{equation}
\end{subequations}
The same result is found for the $k=-1$ case. Therefore, these results confirm our previous analysis, the set of histories is consistent and the probability of finding a singularity is null for the flat and open Friedmann universes filled with radiation.

\subsection{Spherical universe ($k=+1$)} \label{fechado}
For the spherical case, we evaluate the consistency of the families of histories directly by solving the integral in eq. (\ref{consistencia_completaeq}), where we consider the wave function (\ref{funcao_de_onda}). The solution is
\begin{widetext}
\begin{align}
\Braket{\Psi_b |\Psi_s} &\propto \frac{\csc{t_1}\csc{t_1-t_2}}{|\sqrt{\beta+12\imath \tan{(t_1/2)}}|^2} 
\times ( ( \frac{1}{\sqrt{1/|d(t_1,t_2)|}} \fresnelc (\lambda \sqrt{1/|d(t_1,t_2)|}) 
+  \imath \fresnels(\sqrt{1/|d(t_1,t_2)|})\sgn(1/d(t_1,t_2))  ) \nonumber \\
&+ ( \frac{1}{\sqrt{|d(t_1,t_2)|}} \fresnelc (\lambda \sqrt{|d(t_1,t_2)|})-\imath \fresnels(\sqrt{|d(t_1,t_2)|})\sgn(d(t_1,t_2)))) \nonumber \\
&\times( (\sqrt{|d(t_1,t_2)|}(1-2\fresnelc(\lambda \sqrt{1/|d(t_1,t_2)|})+\imath(-1+2 \fresnels(\lambda \sqrt{1/|d(t_1,t_2)|}))\sgn(d(t_1,t_2)) )) \nonumber \\
&+ (\frac{1}{\sqrt{|d(t_1,t_2)|}}(1-2\fresnelc(\lambda \sqrt{|d(t_1,t_2)|})-\imath(-1+2\fresnels(\lambda \sqrt{|d(t_1,t_2)|}))\sgn(d(t_1,t_2)) ))),
\label{consistenciak1}
\end{align}
\end{widetext}
where $d(t_1,t_2):=\tan(\frac{t_1-t_2}{2})$. Eq. (\ref{consistenciak1}) oscillates in the limits $t_1 \rightarrow -\infty$ and $t_2 \rightarrow +\infty$. Hence the family of histories does not decohere and is not possible to assign meaningful probabilities in this case.

%%%%%%%%%%%%%%%%%%%%%%%%%%%

\section{Conclusion}\label{conclusao}

In the present work we have shown that claims asserting that the Wheeler-DeWitt quantization does not eliminate the classical cosmological singularity depend on the quantum theory one is using and on the cosmological model that is being analyzed. In a recent paper \cite{nelsonetal}, it was shown that for a scalar field filled Universe, diferent quantum theories can give different results as long as the resulting Wheeler-DeWitt equation does not reduces to a trivial Schr\"odinger like equation. However, if one considers a FLRW model where the matter content is described by a perfect fluid, where the resulting Wheeler-DeWitt equation reduces to a trivial Schr\"odinger like equation, than, for flat and hyperbolic spatial sections, the conclusions of the Consistent Histories approach calculated in this paper coincide, as it should, with the de Broglie-Bohm calculations made before, namely, that such models never reach a singularity in the two moments considered. As far as we know, this is the first non 
singular 
solution in the Consistent Histories approach concerning quantum cosmology in the WDW quantization. For the $k \rightarrow 1$ case, the existence of singularities can not be answered due to the fact that the family of histories is not consistent and then this approach is silent about this question. Since in this model we expect oscillations between the contracting and expanding phases, the lack of consistency may be related to the specific moments in time we chose.

\begin{acknowledgments}
C.R. Bom was funded by a FAPERJ ``Nota 10'' fellow. G. B. Santos was funded by a PCI grant from MCTi/CNPq of Brazil and N. Pinto-Neto would like to thank CNPq for financial support. We also thank A. Scardua for useful discussions.
\end{acknowledgments}

\appendix
\section{Numerical calculations}
\begin{figure}[!htb]
       \centering  % figura centralizada
       \includegraphics[width=8cm]{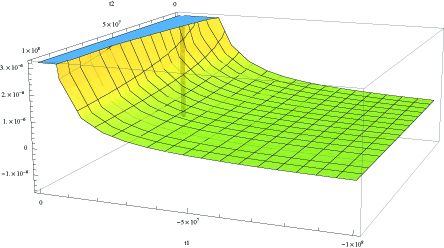}
       \includegraphics[width=8cm]{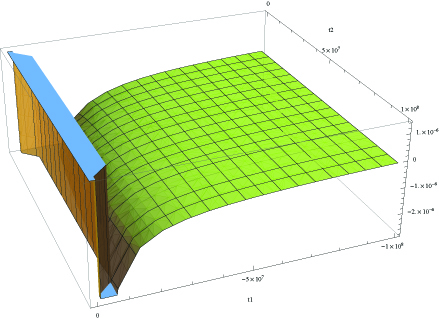}
       \caption{Real and Imaginary parts of $A_0(t_1,t_2)$ respectively where $R_\star=200$, $p=0$ e $\sigma=15$.}
       \label{fig:consistencia_k0}
   \end{figure}

\begin{figure}[!htb]
       \centering  % figura centralizada
       \includegraphics[width=8cm]{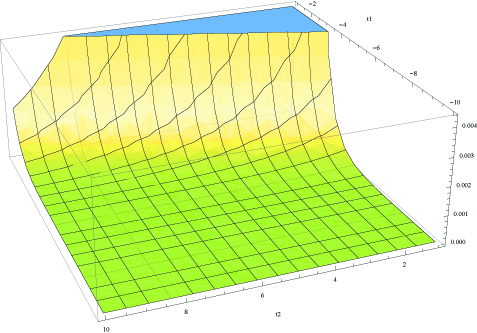}
       \includegraphics[width=8cm]{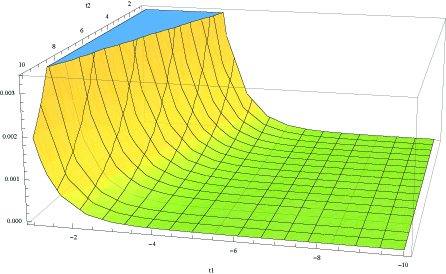}
       \caption{Real and Imaginary parts of $A_{-1}(t_1,t_2)$ respectively, where $R_\star=200$, $p=0$ e $\sigma=15$.}
       \label{fig:consistencia_k-1}
   \end{figure}

According to equations (\ref{consistencia_singular}), (\ref{prob_singularidade}) and (\ref{prob_ricochete}), in order to check the consistency of the family of histories and calculate the probabilities $p_s$, $p_b$, one must evaluate the integrals 
\begin{subequations}
\begin{align}
\Braket{\Psi |P_{\Delta R}(t)|\Psi}&= \Bra{\Psi}U^{\dagger}\int_{\Delta R}dR'\Ket{R'}\Bra{R'}U\Ket{\Psi} \nonumber \\
&=  \int_{\Delta R}dR'|\Psi(R',t)|^2 \quad , 
\label{um_proj}
\end{align}
\begin{align} 
\Braket{\Psi |P^{2}_{\Delta R}P^{1}_{\Delta R}|\Psi} &= \nonumber \\ 
&= \int_{\Delta R}dR \int_{\Delta R}dR'G^{(a)}(R,R',t_2-t_1) \nonumber \\ 
& \quad \times \Psi(R',t_1) \Psi^\star(R,t_1) \quad , 
\label{dois_proj}
\end{align}
\begin{align} 
\Braket{\Psi |P^{1}_{\Delta R}P^{2}_{\Delta R}P^{1}_{\Delta R}|\Psi} &= \int_{\Delta R}dR\int_{\Delta R}dR'\int_{\Delta R}dR'' \nonumber \\
& \times \Psi^\star(R,t_1) G^{(a)}(R,R',t_1-t_2) \nonumber \\
& \times G^{(a)}(R',R'',t_2-t_1)\Psi(R'',t_1). 
\label{tres_proj}
\end{align}
\end{subequations}
For the wave function (\ref{funcao_de_onda}) in the $k \rightarrow 0$ case, from equation (\ref{um_proj}), it follows that
\begin{align} 
\Braket{\Psi |P_{\Delta R^\star}(t)|\Psi}&=  \int^{R^\star}_{0} dR |\Psi(R,t)|^2 \nonumber\\
&= \int^{X'}_{0} dx \frac{2}{\sqrt{\pi}}\exp(-x^2)
\end{align}
where $x = \sqrt{\frac{72}{\sigma^2 t^2 +(6-pt)^2}} R$.
In the considered limits we have
\begin{equation}
\lim_{t\rightarrow\pm\infty} \int^{X'}_{0} dx \frac{2}{\sqrt{\pi}}\exp(-x^2)
= 0.
\end{equation}
Likewise, for the hyperbolic case Eq. (\ref{um_proj}) yields
\begin{align} 
\Braket{\Psi |P_{\Delta R^\star}(t)|\Psi}&= \int^{R^\star}_{0} dR |\Psi(R,t)|^2 \nonumber\\
& =  \int^{Y'}_{0} dy \frac{2}{\sqrt{\pi}}\exp(-y^2)
\end{align}
where $$y =  \frac{\sqrt{72 \sigma}}{\cosh t(\sigma^2 \tanh^2 t+(6-p\tanh t)^2)^{1/2}} R.$$
Also in this case we get
\begin{equation}
\lim_{t\rightarrow\pm\infty} \int^{Y'}_{0} dy \frac{2}{\sqrt{\pi}}\exp(-y^2) = 0.    
\end{equation}
To determine the consistency of the family of histories we evaluate Eq.\,(\ref{tres_proj}) numerically. We present the results for the $k \rightarrow 0$ case in figure (\ref{fig:consistencia_k0}), where $A_0(t_1,t_2):=\Braket{\Psi |P^{2}_{\Delta R}P^{1}_{\Delta R}|\Psi}$ and we have used $R_\star=200$ (of order of hundreds of Planck lengths) and $\sigma=15$. The constant $p$ is set to $0$ since it can be eliminated by a time translation. In figure (\ref{fig:consistencia_k-1}) we present the results for the $k \rightarrow -1$ case, assuming the same set of constants. In both cases we have 
\begin{equation}
\lim_{\substack{t_1\rightarrow -\infty\\ t_2\rightarrow +\infty }} \Braket{\Psi |P^{2}_{\Delta R}P^{1}_{\Delta R}|\Psi} = 0.    
\end{equation}

%\end{widetext}
It follows from eq.(\ref{consistencia_singular}) that the family of histories is consistent for flat and hyperbolic cases. To calculate the probability of a singular universe it also is necessary to solve eq. (\ref{tres_proj}). The solution for this equation for the flat universe is 


\begin{widetext}
\begin{align}
\Braket{\Psi |P^{1}_{\Delta R_\star}P^{2}_{\Delta R_\star}P^{1}_{\Delta R_\star}|\Psi} & = \frac{\sigma R_\star}{\sqrt{2\pi^3}|\beta|^2 t_1(t_1-t_2)} 
& \times \left (12R_\star \sgn(t_1-t_2) + \sqrt{3\pi|t_1-t_2|} \right)^2 \nonumber \\
\times \left| \left(  \fresnelc(\frac{\sqrt{3}\cdot 4R_\star}{\sqrt{\pi|t_1-t_2|}})\sgn(t_1-t_2) \right.\right.
& \left.\left.\quad + \imath \fresnels( \frac{\sqrt{3}\cdot 4R_\star}{\sqrt{\pi|t_1-t_2|}}) \right) \right|^2
\label{tres_proj_k0}, 
\end{align} 
where $\fresnels(y)=\int_{0}^{y}dt \sin{\frac{\pi t^2}{2}}$ e $\fresnelc(y)=\int_{0}^{y}dt \cos{\frac{\pi t^2}{2}}$ are the Fresnel $\sin$ and $\cos$ functions respectively. In the hyperbolic case the solution found is
\begin{align}
\Braket{\Psi |P^{1}_{\Delta R_\star}P^{2}_{\Delta R_\star}P^{1}_{\Delta R_\star}|\Psi} &= \sqrt{3}/2 \lambda \left|\sqrt{\frac{-\imath \cosh(t_1)}{\beta - 12\imath \tanh(t_1/2)}}\right|^2 \cosh(t_1-t_2) 
\times  [\fresnelc{}^2(\lambda \sqrt{1/b(t_1,t_2)}) + \nonumber \\
&+ \frac{2 \fresnelc(\lambda \sqrt{1/b(t_1,t_2)})\fresnelc(\lambda \sqrt{b(t_1,t_2)})}{|b(t_1,t_2)|} + \frac{1}{b^2(t_1,t_2)}(\fresnelc{}^2(\lambda \sqrt{b(t_1,t_2)})\nonumber \\
& + \fresnels {}^2(\lambda \sqrt{b(t_1,t_2)}) )  + \fresnels {}^2(\lambda \sqrt{1/b(t_1,t_2)})\sgn {}^2(1/b(t_1,t_2)) \nonumber \\
&+ \frac{2}{b{}^{3/2}(t_1,t_2)}\fresnels{}^2(\lambda \sqrt{1/b(t_1,t_2)}) \times \fresnels{}^2(\lambda \sqrt{b(t_1,t_2)})\sgn(b^{-1}(t_1,t_2)) \times \sqrt{b(t_1,t_2)}  ]b(t_1,t_2),
\label{tres_proj_k-1}
\end{align}
\end{widetext}
where we defined $b(t_1,t_2):=\tanh(\frac{t_1-t_2}{2})$ and $\lambda:=2R_\star \sqrt{\frac{6}{\pi}}$. Once more for both flat and hyperbolic cases it follows that
\begin{equation} 
\lim_{\substack{t_1\rightarrow -\infty\\ t_2\rightarrow +\infty }}\Braket{\Psi |P^{1}_{\Delta R}P^{2}_{\Delta R}P^{1}_{\Delta R}|\Psi}
=
\left\{
\begin{array}{lcl}   
         0             && t_1\rightarrow +\infty   \\
         0             && t_1\rightarrow -\infty    \\  
         0             && t_2\rightarrow +\infty   \\
         0             && t_2\rightarrow -\infty    \\                     
\end{array}\right.  
\label{tres_proj_k0_limite}
\end{equation}
With the help of equations (\ref{consistencia_singular}), (\ref{prob_singularidade}) and (\ref{prob_ricochete}) we find that 
\begin{subequations}
\begin{equation}
\lim_{\substack{t_1\rightarrow -\infty\\ t_2\rightarrow +\infty }} \Braket{\Psi_b|\Psi_s}=0 ,
\end{equation}
\begin{equation}
\lim_{\substack{t_1\rightarrow -\infty\\ t_2\rightarrow +\infty }} \Braket{\Psi_s|\Psi_s}=0 ,
\end{equation}
\begin{equation}
\lim_{\substack{t_1\rightarrow -\infty\\ t_2\rightarrow +\infty }} \Braket{\Psi_b|\Psi_b}=1 .
\end{equation}
\end{subequations}
The universe never reaches a singularity in those two limits considered for flat and hyperbolic universes.

\end{document}